\documentstyle[12pt,epsfig]{article}

\newcommand{\be}{\begin{equation}}
\newcommand{\ee}{\end{equation}}
\newcommand{\Dlt}{\Delta}

\newcommand{\om}{\omega}

\newcommand{\bS}{{\bf S}}
\newcommand{\bI}{{\bf I}}
\newcommand{\bB}{{\bf B}}
\newcommand{\bt}{\beta}
\newcommand{\al}{\alpha}
\newcommand{\gm}{\gamma}
\newcommand{\Gm}{\Gamma}

\begin{document}

\begin{center}
{\Large{\bf Coherent radiation by molecular magnets} \\ [5mm]
V.I. Yukalov$^{1,2}$ and E.P. Yukalova$^{1,3}$} \\ [3mm]
{\it
$^{1}$Institut f\"ur Theoretische Physik, \\
Freie Universit\"at Berlin, Arnimallee 14, D-14195 Berlin, Germany\\ [2mm]
$^{2}$Bogolubov Laboratory of Theoretical Physics, \\
Joint Institute for Nuclear Research, Dubna 141980, Russia \\ [2mm]
$^{3}$Department of Computational Physics, Laboratory of Information
Technologies, \\
Joint Institute for Nuclear Research, Dubna 141980, Russia}

\end{center}

\vskip 1cm

\begin{abstract}

The possibility of coherent radiation by molecular magnets is investigated. 
It is shown that to realize the coherent radiation, it is necessary to couple 
the considered sample to a resonant electric circuit. A theory for describing 
this phenomenon is developed, based on a realistic microscopic Hamiltonian, 
including the Zeeman terms, single-site anisotropy, and dipole interactions.
The role of hyperfine interactions between molecular and nuclear spins is
studied. Numerical solutions of the spin evolution equations are presented.

\end{abstract}

\vskip 2cm

{\bf PACS}: 

07.57.Hm - Infrared, microwave, and radiowave sources

75.50.Xx - Molecular magnets

76.20.+q - General theory of magnetic resonances and relaxation

\newpage

Molecular magnets is a relatively new class of materials possessing several 
interesting properties (see e.g. reviews [1--3]). One of the important 
features is that the molecules, forming the magnet, can have rather different 
values of spin, between $1/2$ and more than $10$. At low temperatures, the
molecular magnets are bistable and exhibit hysteresis [4] and tunneling 
between up and down orientations [5,6]. Their potential use for quantum 
computation has been proposed [7], and experiments with millimeter-wave 
radiation have been carried out showing that the relaxation rate for 
magnetization reversal and the energy-level populations can be controlled
[8--12]. The magnetization reversal, caused by a linearly varying in time 
external magnetic field, with a fast magnetization avalanche [4], is due 
to the Landau-Zener tunneling [13--16]. The latter is a quantum effect, 
involving no coherence between different molecules. An attempt has been 
made [17] to check if the quantum tunneling can be accompanied by coherent 
electromagnetic radiation. However, the most recent and accurate 
experiment [18] detected no significant radiation.

In the present paper, we suggest a principally different setup for realizing 
coherent radiation from molecular magnets. The key difference is that a magnet
has to be coupled to a resonant electric circuit. Such a setup has proved to 
be effective in achieving coherent radiation by nuclear spins, which was 
described theoretically [19--21] as well as performed experimentally (see 
reviews [22--24]). Molecular magnets, however, have several properties 
essentially distinguishing them from nuclear magnets. The most important 
distinctions are the possibility of possessing a high spin, the existence of 
level splitting, the presence of a single-site anisotropy that can be rather 
large, and the necessity of taking account of the line-width narrowing effect 
due to a strong spin polarization. The aim of this letter is to present a 
generalized theory of coherent spin radiation, including these principal 
features, typical of molecular magnets, and to give numerical solutions of 
the derived generalized equations.

Our consideration is based on the macroscopic Hamiltonian
\be
\label{1}
\hat H = \sum_i \hat H_i + \frac{1}{2}\; \sum_{i\neq j} \hat H_{ij} \; ,
\ee
describing $N$ molecules, each having spin $S$. The first term in Eq. (1)
contains the single-spin Hamiltonian
\be
\label{2}
\hat H_i = -\mu_0 \bB \cdot \bS_i - D(S_i^z)^2 \; ,
\ee
in which $\mu_0=\hbar\gm_S$, with $\gm_S$ being the gyromagnetic ratio,
$\bB$ is the total magnetic field acting on spin $\bS_i$, and $D$ is the
anisotropy parameter. The total magnetic field
\be
\label{3}
\bB = B_0{\bf e}_z + (h_0 + h_1\cos\om t + H){\bf e}_x 
\ee
consists of a longitudinal magnetic field $B_0$, transverse constant field
$h_0$, modelling the level splitting, an alternating field $h_1\cos\om t$,
and the resonator field $H$. The latter is defined by the Kirchhoff 
equation
\be
\label{4}
\frac{dH}{dt} + 2\gm H + \om^2 \int_0^t H(t')\; dt' = 
\gm e_f - 4\pi\eta\; \frac{dm_x}{dt} \; ,
\ee
in which $\gm$ is the circuit damping, $\om$ is the circuit natural frequency,
$e_f=h_2\cos\om t$ is an electromotive force, $\eta$ is a filling factor, and 
$m_x\equiv(\mu_0/V)\sum_i<S_i^x>$ is the average transverse magnetization.
The coupling of a molecular magnet to a resonant electric circuit through 
the Kirchhoff equation (4) is the fundamental point in the proposed setup.
The second term in Hamiltonian (1) corresponds to dipole spin interactions,
with
\be
\label{5}
\hat H_{ij} = \sum_{\al\bt} D_{ij}^{\al\bt} S_i^\al S_j^\bt \; ,
\ee
where $D_{ij}^{\al\bt}$ is the dipolar tensor.

The necessary conditions for the occurrence of coherent spin motion are 
the existence of a well-defined frequency of their rotation $\om_s$ and 
the formation of spin packets of a characteristic size $L_s$, such that 
$kL_s\ll 1$, where $k\equiv\om_s/c$. Then the arising coherent radiation 
happens at the frequency $\om_s$. If the radiation wavelength is larger 
than the system linear size $L$, then the spin-packet length $L_s$, just 
coincides with $L$. This, however, is not compulsory, and $L_s$ can be 
much shorter than $L$. But the condition $kL_s\ll 1$ is necessary for spin 
packets to be formed. Writing down the Heisenberg equations of motion for 
the spin operators, we shall average them over the spin degrees of freedom,
aiming at obtaining the evolution equations for the following averages: The
transition function
\be
\label{6}
u \equiv \frac{1}{SN_s} \; \sum_{i=1}^{N_s} \; < S_i^-> \; ,
\ee
the coherence intensity
\be
\label{7}
w \equiv \frac{1}{S^2N_s(N_s-1)} \; 
\sum_{i\neq j}^{N_s} \; <S_i^+ S_j^-> \; ,
\ee
and the spin polarization
\be
\label{8}
s \equiv \frac{1}{SN_s}\; \sum_{i=1}^{N_s} \; <S_i^z> \; ,
\ee
where $N_s\equiv\rho L_s^3$ is the number of spins in a spin packet, 
$\rho\equiv N/V$ is the average spin density, and $S_i^\pm$ are the ladder 
spin operators.

In each spin system, there are dipole spin interactions, represented by 
Eq. (5). These interactions must be taken into account when considering 
any collective properties of the spin system. The dipole interactions is 
the major cause for the appearance of the transverse relaxation rate. A
peculiarity of molecular magnets is the possibility of having very high
longitudinal spin polarizations. This requires to include in the effective 
transverse relaxation the effect of line-width narrowing [25], which 
results in the effective relaxation  rate
\be
\label{9}
\Gm_2 = (1-s^2)\gm_2 + \gm_2^* \; ,
\ee
where $\gm_2\equiv n_0\rho\mu_0^2\sqrt{S(S+1)}/\hbar$ is the spin dephasing
rate, with $n_0$ being the number of nearest neighbours, and $\gm_2^*$ is 
an inhomogeneous broadening [25,26].

From the evolution equations for the spin variables (6) to (8), it follows
that the frequency of spin rotation is
\be
\label{10}
\om_s = \om_0 -\om_D s\; ,
\ee
in which $\om_0\equiv-\mu_0B_0/\hbar$ is the Zeeman frequency and 
$\om_D\equiv(2S-1)D/\hbar$ is the anisotropy frequency. The second term in 
Eq. (10) depends on time through $s=s(t)$. Therefore to keep the spin 
frequency (10) in resonance with the resonator natural frequency $\om$, one 
has to apply a sufficiently strong external magnetic field $B_0$, such that 
$|\om_0/\om_D|\gg 1$, and the resonance condition $|\Dlt/\om|\ll 1$, 
$(\Dlt\equiv\om -|\om_s|)$ could be realized. In some molecular magnets, 
the single-site anisotropy is rather strong, so that $\om_D$ can reach the 
values of $10^{12}$ s$^{-1}$. Then, to make $|\om_0|$ much larger then 
$\om_D$, one has to invoke magnetic fields $B_0\sim 10-100$ T. Fortunately, 
many molecular magnets possess much weaker anisotropies, for which $B_0\sim 1$
T would be quite sufficient. In addition, achieving nowadays very high 
magnetic fields is not an insuperable obstacle. Among available sources [27], 
there are those allowing to reach the stationary magnetic fields of $45$ T 
(National High Magnetic Field Laboratory, USA) and the pulsed fields up to 
$600$ T (University of Tokyo, Japan). The pulsed fields can be supported 
during the time $10^{-2}$ s up to several seconds, which is perfectly 
sufficient for realizing coherent radiation in a molecular magnet.

In what follows, we assume that there is an external constant magnetic 
field $B_0$, for which the resonance condition $|\Dlt/\om|\ll 1$ is valid, 
that is the resonant circuit can be tuned to the resonance with the spin 
frequency (10). As is mentioned above, to achieve coherent radiation, one 
has to have a well-defined frequency $\om_s$, which requires that the
magnitude of the latter be much larger than any attenuation rate. If so, 
the system possesses two different time scales, which allows us to employ
the scale-separation approach [19--23], whose mathematical foundation is
based on the averaging techniques [28]. The transverse function (6) is fast, 
as compared to the slow variables (7) and (8). The latter are to be treated 
as quasi-invariants, when solving the equation for function (6).

To represent the final formulas in a nice way, we need to introduce several
notations. The spin-resonator coupling is characterized by the dimensionless 
coupling parameter
\be
\label{11}
g \equiv \frac{\gm\gm_0\om_s}{\gm_2(\gm^2+\Dlt^2)} \; ,
\ee
in which $\gm_0\equiv\pi\eta\rho\mu_0^2S/\hbar$ is the natural spin width.
The action of the resonator feedback field comes through the temporal 
coupling function
\be
\label{12}
\al\equiv g\gm_2\left ( 1 - e^{-\gm t}\right ) \; .
\ee
The effective transverse attenuation is given by the collective width
\be
\label{13}
\Gm \equiv \Gm_2 +\gm_3 -\al s \; ,
\ee
in which $\gm_3$ is the inhomogeneous dynamic broadening due to local 
dipole fluctuations [19--23].

For the transverse function (6), we find
$$
u \cong -\; \frac{\nu_0 s}{\om_s -i\Gm} + \frac{(\nu_1+\bt)s}{\Dlt+i\Gm}\;
e^{-i\om t} + \left [ u_0 + \frac{\nu_0s}{\om_s-i\Gm}\; - \;
\frac{(\nu_1+\bt)s}{\Dlt+i\Gm}\right ] e^{-(i\om_s+\Gm)t} \; ,
$$
where $\nu_0\equiv\mu_0h_0/\hbar$, $\nu_1\equiv\mu_0h_1/2\hbar$, 
$\nu_2\equiv\mu_0h_2/2\hbar$, and the function 
$\bt\equiv(\nu_2/2)(1-e^{-\gm t})$ describes the action of an 
electromotive force, if any.

Let us define the effective attenuation
\be
\label{14}
\Gm_3 \equiv \gm_3 + \frac{\nu_0^2\Gm}{\om_s^2 +\Gm^2} \; - \;
\frac{\nu_0(\nu_1+\bt)\Gm}{\om_s^2+\Gm^2}\; e^{-\Gm t} + 
\frac{(\nu_1+\bt)^2\Gm}{\Dlt^2+\Gm^2}\; \left ( 
1 - e^{-\Gm t} \right ) \; .
\ee
Then the final evolution equations for the slow functions (7) and (8) can 
be represented in the form
\be
\label{15}
\frac{dw}{dt} = - 2(\Gm_2 -\al s) w + 2\Gm_3 s^2 \; , \qquad
\frac{ds}{dt} = -\al w - \Gm_3 s - \Gm_1(s -\zeta) \; ,
\ee
in which $\zeta$ is a stationary spin polarization and $\Gm_1=\gm_1+\gm_1^*$ 
is the sum of the spin-lattice attenuation $\gm_1$ and a pumping rate 
$\gm_1^*$, if the sample is subject to a nonresonant pumping procedure.

The evolution equations (15) include all important attenuation rates that 
can influence spin dynamics. In general, there exists also one more 
relaxation parameter, the radiation rate $\gm_r=2\mu_0^2 Sk^3N_s/3\hbar$ 
due to spin interactions through the common radiation field. However this
rate, as has been noticed yet by Bloembergen [29], does not influence the
dynamics of a macroscopic spin sample. This is evident from the ratio
$\gm_r/\gm_2\approx 0.1(kL_s)^3\ll 1$, which shows that $\gm_r$ is negligible
as compared to the line width formed by the dipole interactions. Such a 
situation is drastically different from atomic systems [22], in which both 
the dephasing rate $\gm_2$ as well as the collective radiation rate $\gm_r$ 
are caused by the same reason, by atomic interactions through the photon 
exchange, so that $\gm_r/\gm_2\sim N_c\gg 1$, with $N_c$ being the number
of correlated atoms in a wave packet. But in spin systems, $\gm_r$ and 
$\gm_2$ are due to different origins. As is thoroughly explained in review
[24], in resonatorless spin systems interacting through dipole forces, the
effect of pure spin superradiance is principally inacheivable. This is why, 
for organizing coherent radiation by a magnet, one has to couple the latter 
to a resonant circuit.

Let us investigate the spin dynamics, described by Eqs. (15), for a molecular 
magnet, such as Mn$_{12}$ or Fe$_8$, whose molecules possess spin $S=10$.
Then $\om_D\sim 10^{12}$ s$^{-1}$. At low temperatures, below the blocking
temperature of about $1$ K, the sample can be polarized, keeping its 
polarization for rather long times of order $T_1\sim 10^5-10^7$ s, hence 
$\gm_1\equiv T_1^{-1}$ is very small, $\gm_1\sim 10^{-7}-10^{-5}$ s$^{-1}$.
The density of molecules in a magnet is $\rho\sim 10^{20}$ cm$^{-3}$. Dipole
interactions are quite strong, yielding $\gm_2\sim 10^{10}$ s$^{-1}$. The
tunnel splitting is small, with $\nu_0\sim 10^4$ s$^{-1}$, which is much
smaller than $\om_D$.

Assume that the sample is polarized, with the spin polarization up, 
$s_0\equiv s(0)>0$. Then it is placed in an external magnetic field $B_0$,
such that the Zeeman frequency be positive, $\om_0>0$, hence $\mu_0B_0<0$.
Since $\mu_0=-2\mu_B$, where $\mu_B$ is the Bohr magneton, we have 
$\mu_B B_0>0$. This means that the magnetic field is directed up, which 
implies that the magnet is prepared in a nonequilibrium state. The relaxation
from this state is governed by Eqs. (15). We have analysed these equations
numerically for a wide range of parameters. Our main concern has been to 
study collective effects without imposing transverse fields. So, we present
here the results of our calculations, when $\nu_1=\bt=0$, as a consequence 
of which $\Gm_3\approx\gm_3$. We assume that there is no pumping, so that 
$\gm_1^*=0$, and $\Gm_1=\gm_1\ll\gm_2$. The coupling parameter (11) is of 
the order of the resonator quality factor, $g\sim Q$. We have investigated 
the behaviour of the solutions to Eqs. (15) for the parameters $\gm_2^*$, 
$\gm_3$, $\gm$, and $g$ varied in a very wide diapason, up to three orders 
of magnitude, and for different initial conditions $w_0$ and $s_0$. For a 
strong spin-resonator coupling $g\gg 1$, all solutions, under fixed initial 
conditions, are qualitatively the same. Therefore, it is sufficient to 
demonstrate typical classes of solutions for a fixed set of the parameters 
$\gm_2^*$, $\gm_3$, $\gm$, and $g$, but for distinct initial conditions. It 
is possible to distinguish three such qualitatively different classes of 
solutions, for each of which the relaxation time is much shorter than 
$T_2\equiv 1/\gm_2$.

Figure 1 demonstrates the {\it pure spin superradiance}, starting from a 
high initial spin polarization $s_0=1$, and without any initial coherence 
imposed on the sample, when $w_0=0$. This means that superradiance develops 
as a purely self-organized process. Figure 2 shows the process of {\it 
triggered spin superradiance}, when the initial spin polarization $s_0=0.7$ 
is yet high, an essential self-organization is present, but the relaxation 
is triggered by an initially induced coherence $w_0=0.51$. And Fig. 3 
presents the effect of {\it collective spin induction}, in which the initial 
spin polarization $s_0=0.1$ is low, playing not so important role, while the 
relaxation is induced by a high initial coherence $w_0=0.99$. In all the 
cases, the relaxation time, much shorter than $T_2$, is due to collective 
effects caused by the resonator feedback field, with a strong spin-resonator 
coupling $g\gg 1$. If the latter is weak, $g\leq 1$, then collective coherent 
effects do not develop, and the relaxation time is of the order of $T_2$.
Recall that the initial coherence function $w_0=|u_0|^2$ is nothing but the
amplitude of the transverse spin squared. Therefore to create a nonzero $w_0$ 
it is sufficient to get a nonzero transverse spin polarization $u_0$, which
can be obtained by imposing at the initial time a sufficiently large 
transverse pulsed field.

In molecular magnets, in addition to dipolar fields, there also exist 
hyperfine fields due to interactions between molecular and nuclear spins. 
The influence of the latter can be taken into account by generalizing the 
Hamiltonian (1) to the form $\hat H=\hat H_S + \hat H_I +\hat H_{IS}$, in 
which $\hat H_S$ is the same Hamiltonian (1) due to $S$-spins, the second 
term is the Hamiltonian of nuclear spins ${\bf I}$,
$$
\hat H_I = -\mu_I \sum_i \bB\cdot\bI_i + \frac{1}{2}\;
\sum_{i\neq j}\sum_{\al\bt} D_{ijI}^{\al\bt} I_i^\al I_j^\bt \; ,
$$
where $D_{ijI}^{\al\bt}$ is a dipolar tensor for nuclear spin interactions,
and the hyperfine interactions are described by the Hamiltonian
$$
\hat H_{IS} = \sum_i A\bS_i \cdot\bI_i + \sum_{\i\neq j}\sum_{\al\bt}
A_{ij}^{\al\bt} S_i^\al I_j^\bt \; ,
$$
where $A$ is the intensity of single-site hyperfine interactions and 
$A_{ij}^{\al\bt}$ is the dipolar tensor of hyperfine interactions at different 
sites. The following consideration can be done in the same way as earlier, 
by employing the scale separation approach [19--23]. Hyperfine interactions 
result in the appearance of an additional line broadening of the order of
$10^7-10^8$ s$^{-1}$, which is much smaller than the dipolar line width 
$\gm_2\sim 10^{10}$ s$^{-1}$.

The most nontrivial effect caused by the presence of the hyperfine 
interactions is that the ensemble of nuclear spins, being also coupled to an 
electric circuit, serves as an additional resonator for molecular spins. This 
changes the effective coupling of molecular spins with the resonator 
circuit from Eq. (11) to the value
$$
g =\frac{\gm\gm_0\om_S}{\gm_2(\gm^2+\Dlt^2)}\left ( 1 +
\frac{\rho\mu_I As_I I}{\rho\mu_0\hbar\om_I}\right ) \; ,
$$
where $\om_S\equiv\om_0-\om_D s+(A/\hbar)s_I S$ and
$\om_I\equiv\om_{0I}+(A/\hbar)sI$, with $\om_{0I}\equiv-\mu_IB_0/\hbar$ 
being the nuclear Zeeman frequency, and $s_I$ is an average nuclear spin 
polarization defined analogously to Eq. (8). Estimates show that the change 
in the coupling parameter $g$, owing to the presence of nuclear spins, is 
very small, of the order of $10^{-5}$. Thus, hyperfine interactions do not 
much influence the collective effects of molecular spins.

The intensity of radiation from each spin packet is
$I(t)=(2/3c^3)\mu_0^2S^2\om^4 N_s^2w(t)$, which, depending on the parameters, 
can reach very high values of many Watts. This microwave radiation can be used
for creating spin masers [3]. It is also feasible to organize the regime of 
punctuated spin superradiance [30], which can be employed for information 
processing.

\vskip 5mm

{\bf Acknowledgments}

\vskip 2mm

Financial support from the German Research Foundation (grant Be 142/72-1)
is appreciated. One of the authors (V.I.Y.) is grateful to the German
Research Foundation for the Mercator Professorship.

\newpage

\begin{figure}
\centerline{\psfig{file=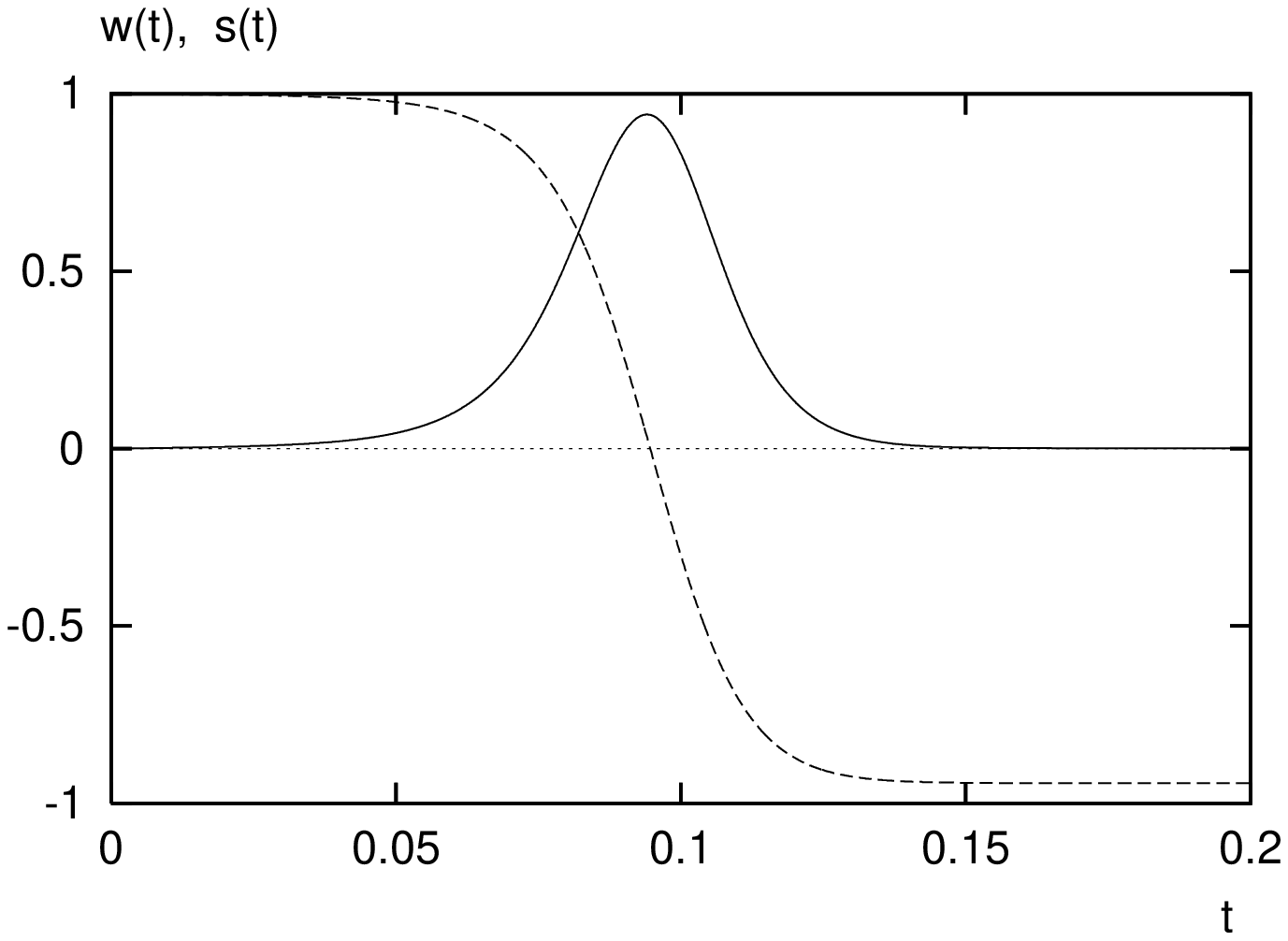,angle=270}}
\caption{Pure spin superradiance. Typical behaviour of the 
coherence intensity $w(t)$ (solid line) and of the spin polarization 
$s(t)$ (dashed line) as functions of time (in units of $T_2$) for the 
following attenuation parameters (in units of $\gm_2$): $\gm_2^*=1$, 
$\gm_3=0.1$, and $\gm=10$. The spin-resonator coupling is $g=100$. The 
initial conditions are $w_0=0$ and $s_0=1$}
\label{Fig.1} 
\end{figure}

\begin{figure}
\centerline{\psfig{file=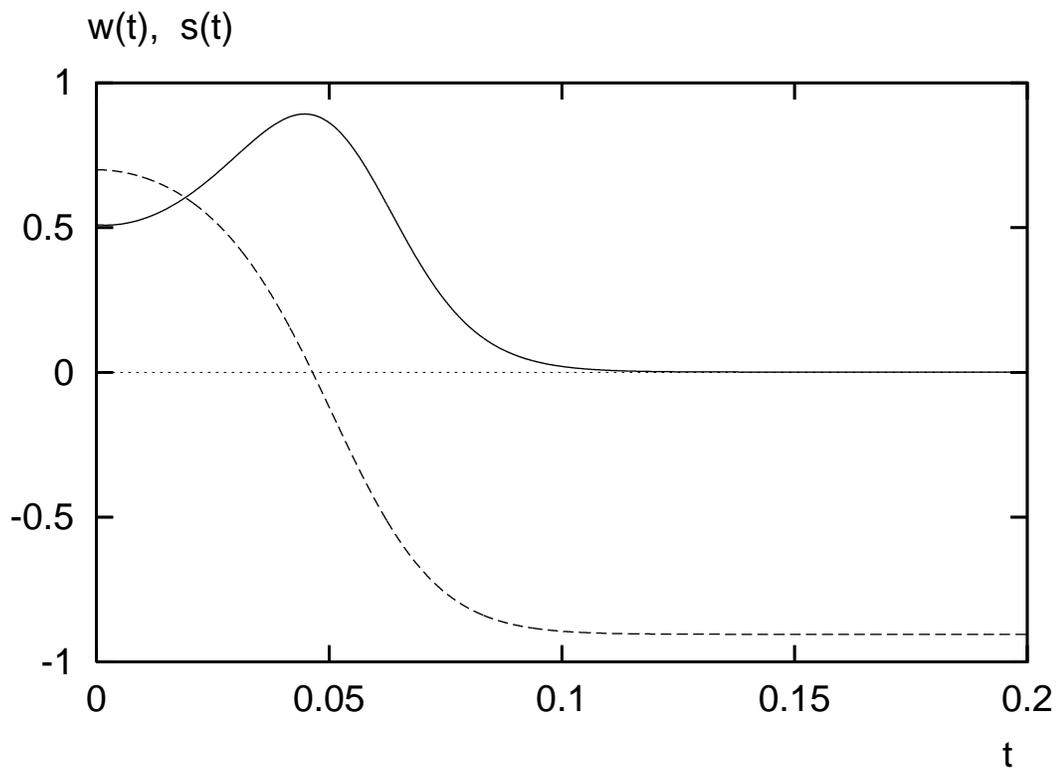,angle=270}}
\caption{Triggered spin superradiance. The coherence intensity $w(t)$
(solid line) and spin polarization $s(t)$ (dashed line) versus time (in
units of $T_2$) for the same parameters as in Fig. 1, but for the 
initial conditions $w_0=0.51$ and $s_0=0.7$}
\label{Fig.2}
\end{figure}

\begin{figure}
\centerline{\psfig{file=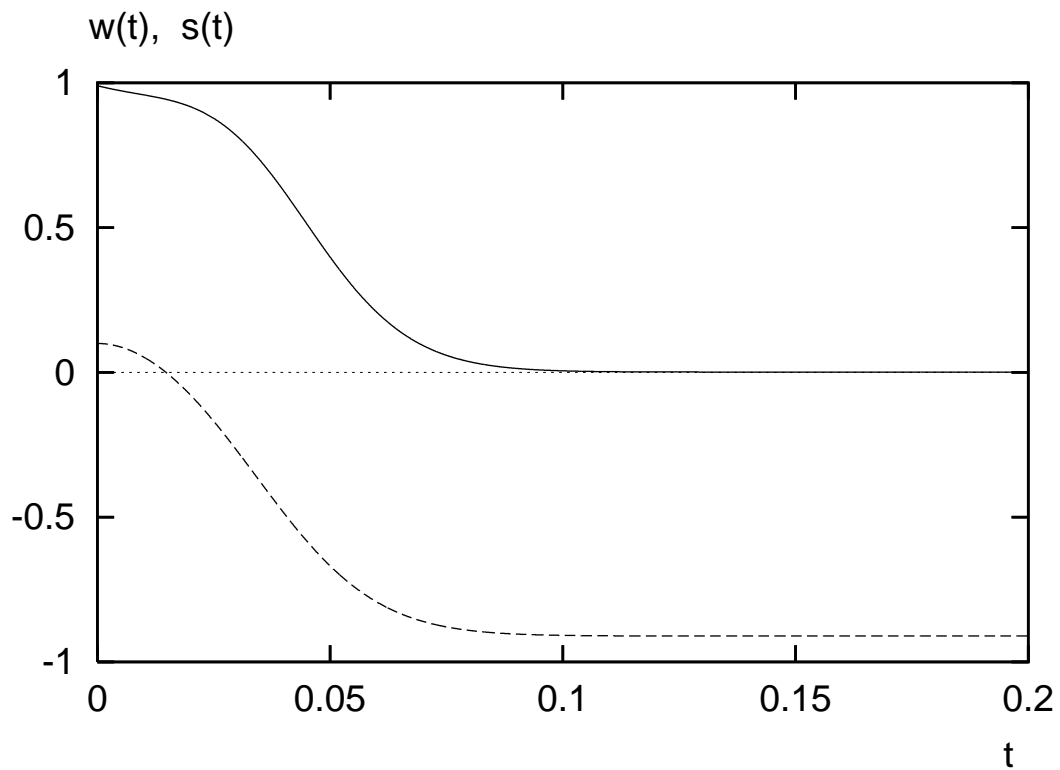,angle=270}}
\caption{Collective spin induction. The coherence intensity $w(t)$
(solid line) and spin polarization $s(t)$ (dashed line) as functions of
time (in units of $T_2$) for the same parameters as in Fig. 1, but for 
the initial conditions $w_0=0.99$ and $s_0=0.1$}
\label{Fig.3} 
\end{figure}

\end{document}